\documentclass[showpacs,byrevtex,groupedaddress,onecolumn,prd]{revtex4}%
\usepackage{amsfonts}
\usepackage{graphicx}
\usepackage{amsmath}
\usepackage{amssymb}%
\providecommand{\U}[1]{\protect\rule{.1in}{.1in}}

\begin{document}
\title{Appearance of Gauge Fields and Forces beyond the adiabatic approximation}
\author{Pierre Gosselin$^{1}$ and Herv\'{e} Mohrbach$^{2}$}
\affiliation{$^{1}$Institut Fourier, UMR 5582 CNRS-UJF, UFR de Math\'{e}matiques,
Universit\'{e} Grenoble I, BP74, 38402 Saint Martin d'H\`{e}res, Cedex, France }
\affiliation{$^{2}$Laboratoire de Physique Mol\'{e}culaire et des Collisions, ICPMB-FR CNRS
2843, Universit\'{e} Paul Verlaine-Metz, 57078 Metz Cedex 3, France}

\begin{abstract}
We investigate the origin of quantum geometric phases, gauge fields and forces
beyond the adiabatic regime. In particular, we extend the notions of geometric
magnetic and electric forces discovered in studies of the Born-Oppenheimer
approximation to arbitrary quantum systems described by matrix valued quantum
Hamiltonians. The results are illustrated by several physical relevant examples.

\end{abstract}
\maketitle

\section{Introduction}

A physical system can never be considered as completely isolated from the rest
of the universe. For a slow (adiabatic) cyclic variation of its environment,
the wave function of the quantum system gets an additional geometric phase
factor, known as the Berry phase \cite{BERRY1}. In fact, the driving
environment, the 'heavy' or 'slow' system, is also subject to back reaction
from the 'light' or 'fast' system. In the context of the Born-Oppenheimer (BO)
theory of molecules, the back reaction of the light system leads to the
appearance of a gauge field in the effective Hamiltonian for the slow one (the
environment) \cite{MEAD}\cite{MOODY}\cite{ZYGELMAN}. The gauge field consists
of a vector and a scalar potential and turns out to depend on a quantum
geometric tensor \cite{BERRY2}. It can both induce interference phenomena and
modify the dynamics through geometric Lorentz and electric forces
\cite{BERRY3}\cite{STERN}. The first measurements of geometric Berry forces
were done, on the one hand in the coordinate space for the evolution of
trapped particles \cite{MULLER} and on the other hand in the momentum space
for the evolution of relativistic particles \cite{BLIOKH4} (see also the
comment in \cite{NORI} which contains related discussion on the geometric
forces and fast-slow motion decoupling in physical systems).

Many physical systems, as discussed in this paper, display a separation of
scales in terms of slow and fast degrees and thus share\ a very similar
mathematical structure with molecular systems. Actually all these systems have
a space-time evolution governed by a multicomponent Schr\"{o}dinger-like
equation, whose Hamiltonian is a matrix valued operator. It is therefore the
purpose of the present note to investigate the origin of quantum gauge fields
and forces in a more general context than the BO theory, by considering the
diagonalization of an arbitrary matrix valued quantum Hamiltonian. To be
precise, by diagonalization we mean the derivation of an effective in-band
Hamiltonian made of block-diagonal energy subspaces. For that purpose we use
the results of a method developed recently \cite{PIERRE1}. This approach,
based on a new differential calculus on a non-commutative space where $\hbar$
plays the role of running parameter, leads to an in-band energy operator that
can be obtained systematically up to arbitrary order in $\hbar.$ Note that
there exists other totally different methods of diagonalization in a formal
series expansion in $\hbar$ which uses symbols of operators via Weyl
calculus\textbf{ }\cite{LITTLEJOHN}\cite{TEUFFEL}\textbf{. }Particularly
important for our purpose, it has been possible, for an arbitrary Hamiltonian
$H(\mathbf{K},\mathbf{Q})$ with the canonical coordinates and momentum
$\left[  Q_{i},K_{j}\right]  =i\hbar\delta_{ij},$ to obtain the corresponding
diagonal representation $\varepsilon\left(  \mathbf{k,q}\right)  $ to order
$\hbar^{2}$, in terms of non-canonical coordinates and momentum ($\mathbf{k,q}%
$\textbf{)} defined later and commutators between gauge fields. The method is
quite involved for an arbitrary Hamiltonian, but simplifies greatly for
systems whose Hamiltonian has the simple form $H=T(\mathbf{K})+V(\mathbf{Q})$.

This kind of Hamiltonian that we are considering in this paper allows us to
discuss how geometric gauge fields and geometric forces arise at order
$\hbar^{2}$ in physical situations as various as Dirac and Bloch electrons in
electric fields or Born-Oppenheimer theory. Note that first order $\hbar$
corrections (semi-classical) were first treated for Bloch electrons in
\cite{NIU} and for Dirac electrons in \cite{ALAIN}\cite{BLIOKH3} (see
\cite{NIU2} for a review and also \cite{BLIOKH2} which contains an overview of
the first-order Berry-phase effects and forces in 4-D space times).

Our approach reveals the appearance at order $\hbar^{2}$ of a scalar gauge
potential expressed in terms of two tensors. One is the quantum metric tensor
\cite{BERRY2}\cite{PROVOST}, and the other one is a new tensor generalizing an
additional term found in \cite{LITTLEJOHN} for the Born-Oppenheimer case.
Another very important consequence of the Hamiltonian diagonalization is the
appearance of gauge invariant intraband coordinates. The advantage of using
these coordinates is that the diagonal Hamiltonian is also gauge invariant.
Moreover, these coordinates fulfill a non-commutative algebra which strongly
affects the dynamics through a Lorentz term and the gradient of a new scalar
potential, generalizing thus the dynamics of the Born-Oppenheimer theory.

\section{Hamiltonian diagonalization.}

Consider the Schr\"{o}dinger equation%
\begin{equation}
i\hbar\frac{\partial\left\vert \Psi\right\rangle }{\partial t}=H\left\vert
\Psi\right\rangle \label{S1}%
\end{equation}
where for the sake of completeness the Hamiltonian is supposed time dependent
and of the form
\begin{equation}
H=T(\mathbf{K,}t)+V(\mathbf{Q,}t) \label{HTV}%
\end{equation}
where it is assumed that $T(\mathbf{K},t)$ is a matrix valued operator while
$V(\mathbf{Q},t)$ is a scalar valued one. We aim to diagonalize the full
differential operator $D=H-K_{0}$ where $K_{0}\equiv i\hbar\partial/\partial
t$ is the conjugate operator of the time $\left[  K_{0},t\right]  =i\hbar.$

The mathematical difficulty in performing the diagonalization of $D$ comes
from the intricate entanglement of noncommuting operators due to the canonical
relation $\left[  Q_{i},K_{j}\right]  =i\hbar\delta_{ij}.$ In \cite{PIERRE1}
starting with a very general but time independent $H(\mathbf{K,Q})$ and by
considering $\hbar$ as a running parameter, we related the in-band Hamiltonian
$UHU^{+}=\varepsilon\left(  \mathbf{X}\right)  $ and the unitary transforming
matrix $U\left(  \mathbf{X}\right)  $ (where $\mathbf{X}\equiv(\mathbf{Q,K})$)
to their classical expressions through integro-differential operators, i.e.
$\varepsilon\left(  \mathbf{X}\right)  =\widehat{O}\left(  \varepsilon
(\widetilde{\mathbf{X}})\right)  $ and $U\left(  \mathbf{X}\right)
=\widehat{N}\left(  U(\widetilde{\mathbf{X}})\right)  $, where in the matrices
$\varepsilon(\widetilde{\mathbf{X}})$ and $U(\widetilde{\mathbf{X}}),$ the
dynamical operators $\mathbf{X}$ are replaced by classical commuting variables
$\widetilde{\mathbf{X}}=(\widetilde{\mathbf{Q}},\widetilde{\mathbf{K}}).$

The only requirement of the method is therefore the knowledge of
$U(\widetilde{\mathbf{X}})$ which gives the diagonal form $\varepsilon
(\widetilde{\mathbf{X}}).$ Generally, these equations do not allow to find
directly $\varepsilon\left(  \mathbf{X}\right)  $, $U\left(  \mathbf{X}%
\right)  $, however, they allow us to produce the solution recursively in a
series expansion in $\hbar.$ With this assumption that both $\varepsilon$ and
$U$ can be expanded in power series of $\hbar,$ we determined in
\cite{PIERRE1}, the explicit diagonalization of an arbitrary Hamiltonian to
order $\hbar^{2}$. The expression of the effective $n$-th in-band energy
$\varepsilon_{n}$ greatly simplifies for Hamiltonian given by Eq. $\left(
\ref{HTV}\right)  $ and the result is given by Eq. (\ref{epsilonh2}). But, for
this simplified problem, it is interesting to give the principal steps leading
to this result. In the first step, assuming that the unitary matrix
$U_{0}(\widetilde{\mathbf{K}}\mathbf{,}t)$ diagonalizing $T(\widetilde
{\mathbf{K}}\mathbf{,}t)$ only, that is $U_{0}TU_{0}^{+}=\varepsilon
_{0}(\widetilde{\mathbf{K}}\mathbf{,}t),$ is known ($\varepsilon
_{0}(\widetilde{\mathbf{K}}\mathbf{,}t)$ is the matrix of the eigenvalues of
$T(\widetilde{\mathbf{K}}\mathbf{,}t)$), we obtain the semiclassical expression%

\[
U_{0}DU_{0}^{+}=\varepsilon_{0}(\mathbf{K,}t)+V(\mathbf{Q+\hbar A,}%
t)-K_{0}+\hbar A^{0}%
\]
with $\mathbf{A(K,}t\mathbf{)}=iU_{0}\nabla_{\mathbf{K}}U_{0}^{+}$ and
$A^{0}(\mathbf{K,}t)=-iU_{0}\frac{\partial}{\partial t}U_{0}^{+}$ two non
diagonal matrices. The diagonalization at the next order ($\hbar^{2}$) is done
by an unitary transformation matrix $U=U_{0}+\hbar U_{1}U_{0}$. The general
method of \cite{PIERRE1} gives an explicit procedure to determine the
antihermitian matrix $(U_{1})_{mn}=\frac{\left(  1-\delta_{mn}\right)
}{\varepsilon_{m}\left(  t\right)  -\varepsilon_{n}\left(  t\right)
}(\mathbf{A)}_{mn}.\nabla V,$ which removes the off-diagonal elements of
$\mathbf{A}$ and $A^{0},$ so that Eq. $\left(  \ref{S1}\right)  $\ becomes
$i\hbar\frac{\partial\left\vert \Psi^{\prime}\right\rangle }{\partial
t}=\varepsilon\left\vert \Psi^{\prime}\right\rangle $ with $\left\vert
\Psi^{\prime}\right\rangle =U\left\vert \Psi\right\rangle $ with the diagonal
energy matrix $\varepsilon$ whose elements $\varepsilon_{n}$ to order
$\hbar^{2}$ are%

\begin{equation}
\varepsilon_{n}\left(  \mathbf{K,q}_{n},t\right)  =\varepsilon_{0,n}\left(
\mathbf{K,}t\right)  +V\left(  \mathbf{q}_{n},t\right)  +\hbar A_{n}^{0}%
+\hbar^{2}\Phi_{n} \label{epsilonh2}%
\end{equation}
where the geometric scalar potential is%

\begin{equation}
\Phi_{n}(\mathbf{\mathbf{Q,K}},t\mathbf{\mathbf{)}}=\frac{G_{n}^{ij}}%
{2}\mathbf{\partial}_{i}\mathbf{\partial}_{j}V\mathbf{+}M_{n}^{ij}%
\mathbf{\partial}_{i}V\partial_{j}V\mathbf{+}\left(  M_{n}^{0i}+M_{n}%
^{i0}\right)  \partial_{i}V+M_{n}^{00} \label{scalar}%
\end{equation}
with two gauge invariant tensors $G_{n}^{ij}$ and $M_{n}^{\mu\nu}$ defined as
\begin{equation}
G_{n}^{ij}(\mathbf{K,}t)=\frac{1}{2}\sum\nolimits_{m\neq n}\left(
(A^{i})_{nm}(A^{j})_{mn}+h.c.\right)  \label{metric}%
\end{equation}
and
\begin{equation}
M_{n}^{\mu\nu}(\mathbf{\mathbf{K,}}t)=\frac{1}{2}\sum\nolimits_{m\neq n}%
(\frac{(A^{\mu})_{nm}(A^{\nu})_{mn}}{\varepsilon_{0,m}-\varepsilon_{0,n}%
}+h.c.) \label{T}%
\end{equation}
Indices $\mu$ corresponding to $\mu=0,1,2,3,$ such that $\mu=0$ is the
temporal variable and $\mu=i=1,2,3$ the spatial ones. The tensor $G_{n}^{ij}$
is known as the quantum metric tensor \cite{BERRY2}\cite{PROVOST} and
$M_{n}^{\mu\nu}$ is a new tensor generalizing an additional term found in
\cite{LITTLEJOHN} for the Born-Oppenheimer theory.

In Eq. $\left(  \ref{epsilonh2}\right)  $ the operator\textbf{ }%
$\mathbf{q}\equiv\mathbf{Q+\hbar A}$ which is non diagonal has been replaced
after application of $U$ by the intraband coordinate\textbf{\ }$\mathbf{q}%
_{n}=\mathbf{Q}+\hbar\mathbf{a}_{n}$\textbf{\ }where\textbf{\ }$\mathbf{a}%
_{n}$\textbf{\ }is a gauge connection usually called the Berry connection
defined as the projection\textbf{ }of $\mathbf{A}\equiv iU_{0}\nabla
_{\mathbf{K}}U_{0}^{+}$ on the $n$-th eigenstate\textbf{\ }$\mathbf{a}%
_{n}(\mathbf{K},t)\equiv(\mathbf{A})_{nn}=i\left\langle n\right.  \left\vert
\nabla_{\mathbf{K}}n\right\rangle $.\textbf{\ }Here\textbf{\ }$\left\vert
n\right\rangle $\textbf{\ }are the eigenstates of the non-diagonal part
of\textbf{\ }$H,$\textbf{\ }i.e.,\textbf{\ }$T(\mathbf{K},t)\left\vert
n\right\rangle =\varepsilon_{0,n}(\mathbf{K},t)\left\vert n\right\rangle
$\textbf{. }In the same manner $a_{n}^{0}\equiv(A^{0})_{nn}=-i\left\langle
n\right.  \left\vert \overset{\cdot}{n}\right\rangle $ is the scalar gauge
potential. Note that even though it is formally indifferent to express\textbf{
}$\Phi_{n}$ in terms of $\mathbf{Q}$ or $\mathbf{q}_{n}$ the difference being
of higher order in $\hbar$, the introduction of the non-canonical
coordinate\textbf{\ }$\mathbf{q}_{n}$\textbf{\ }is\textbf{ }essential to
maintain the gauge invariance of the Hamiltonian. The formal similarity with
gauge theories is evident as we can define a Berry curvature the $n$-th
eigenstate (which can be degenerate)\textbf{\ }%
\begin{equation}
\Theta_{n}^{ij}(\mathbf{K})=\frac{\partial a_{n}^{j}}{\partial K_{i}}%
-\frac{\partial a_{n}^{i}}{\partial K_{j}}+\left[  a_{n}^{i},a_{n}^{j}\right]
\label{courbure}%
\end{equation}
leading to the following non-canonical commutation relations
\begin{equation}
\left[  q_{n}^{i},q_{n}^{j}\right]  =i\hbar^{2}\Theta_{n}^{ij} \label{NC}%
\end{equation}
As usual, the curvature leads to introduce a magnetic-type vector field
$\Theta_{n}(\mathbf{K},t)$ whose components are defined as $\Theta_{n}%
^{i}=\varepsilon_{jk}^{i}\Theta_{n}^{jk}/2$ and in the same manner from the
temporal component $a_{n}^{0}(\mathbf{K},t)$ an electric-type field is defined as%

\begin{equation}
\mathbf{E}_{n}(\mathbf{K,}t)\mathbf{=-}\frac{\partial\mathbf{a}_{n}}{\partial
t}-\nabla a_{n}^{0} \label{electric}%
\end{equation}
Note that extension of the first-order ($\hbar$) Berry adiabatic formalism to
the 4D space-time evolution with the \textquotedblleft
magnetic\textquotedblright\ and \textquotedblleft electric\textquotedblright%
\ geometric fields and the corresponding Hamiltonian approach has been made in
\cite{BLIOKH3}\cite{BLIOKH2}.

For certain systems, like spinning particles in ferromagnets, the
electric-type field leads to spin motive forces (Faraday law of spin)
\cite{BARNES}. Here the motive force is given by
\[
\xi=-\oint\nolimits_{C}\mathbf{E}_{n}\mathbf{.}d\mathbf{K=}\frac{\partial
}{\partial t}\int_{S}\Theta_{n}\mathbf{\cdot}d\mathbf{S}_{K}=\frac
{\partial\phi}{\partial t}%
\]
where $C$ is a closed curve in $\mathbf{K}$ space, and $S$ the surface
delimited by $C$ ($d\mathbf{S}_{K}$ being the infinitesimal surface vector
orthogonal to $S$), and $\phi$ the magnetic type flux through $S$. The motive
force is independent of the charge of the particle and requires only a time
dependent Berry curvature.

The Heisenberg equations of motion $\overset{\cdot}{\mathbf{q}}_{n}%
=-i\hbar\left[  \mathbf{q}_{n},\varepsilon_{n}\right]  +\frac{\partial
\mathbf{q}_{n}}{\partial t}$ et $\overset{\cdot}{\mathbf{K}}=-i\hbar\left[
\mathbf{K},\varepsilon_{n}\right]  $ to the second order in $\hbar$ are
\begin{equation}
\overset{\cdot}{\mathbf{q}}_{n}=\nabla_{\mathbf{K}}\varepsilon_{0,n}%
-\hbar\mathbf{E}_{n}-\frac{\hbar}{2}(\overset{\cdot}{\mathbf{K}}%
\times\mathbf{\Theta}_{n}-\mathbf{\Theta}_{n}\times\overset{\cdot}{\mathbf{K}%
})+\hbar^{2}\nabla_{\mathbf{K}}\Phi_{n},\text{ \ \ \ }\overset{\cdot
}{\mathbf{K}}=-\nabla_{\mathbf{q}_{n}}V \label{EQMVT}%
\end{equation}
The dynamics of the intraband operators leads directly to a Lorentz-type term.
The scalar potential is a consequence of transitions between eigenstates
$\left\vert n\right\rangle $ of $T$ and impacts the dynamics through its
gradient. Working with the non-canonical coordinates is a short-cut to
determine the dynamics of a system prepared in a second order in $\hbar$
eigenstate of the diagonalized Hamiltonian Eq. $\left(  \ref{epsilonh2}%
\right)  $. This state will evolve in the same energy subspace $n$, as there
are no transitions between these eigenlevels as far as we can neglect higher
contributions in the expansion in $\hbar$. In comparison, the equations of
motion derived from the Hamiltonian $H$ do not seem to include a Lorentz
force, and the determination of the "eigendynamics" can be a very difficult to
achieve. An appealing example is given in \cite{BERRY3}\cite{STERN} where the
"exact" slow motion of a massive neutral particle coupled to a spin is
compared with the Born-Oppenheimer theory.\textbf{\ }

We underline that the expansion in $\hbar$ breaks down in regions of mode
conversion where $\varepsilon_{0,m}-\varepsilon_{0,n}<<\hbar$ or for large
values of\ $\left\langle m\right.  \left\vert \partial_{\mu}n\right\rangle
.$\ In a mode conversion region, one can easily generalize the diagonalization
of $H$\ to a block-diagonalization where transitions between states inside the
block are allowed (off-diagonal elements), but there are no transitions
between different blocks. The expansion in $\hbar$ is consistent with the
adiabatic approximation (see \cite{TEUFFEL} and related discussion in
\cite{BLIOKH2}). Indeed, the semiclassical order ($\hbar$) where the potential
$\Phi_{n}$ is absent corresponds to a situation of no-transitions between the
states $\left\vert n\right\rangle $ (or between states in different blocks)
whereas the expansion to second order in $\hbar$ corresponds to post-adiabatic
corrections \cite{BERRY4} where there is no transition of higher order between
eigenlevels and eigenstates of the diagonalized Hamiltonian Eq. $\left(
\ref{epsilonh2}\right)  $.

The virtue of our approach is its full generality which sheds a new light and
provides an unified description of several phenomena which were considered
case by case. We are now going to illustrate the results by several physical examples.

\section{Born-Oppenheimer approximation.\textit{ }}

Consider the following Hamiltonian describing a fast system in interaction
with an external environment
\begin{equation}
H=\frac{1}{2}B_{ij}P^{i}P^{j}+\frac{p^{2}}{2m}+\varphi(\mathbf{R,r})
\end{equation}
where the fast system is described by a set of dynamical variables
$(\mathbf{r,p})$ (not to be confused with the non-canonical coordinate and
momentum operators) and the slow one (the environment) by coordinates
$(\mathbf{R,P}).$ As in \cite{BERRY2} we consider a general kinetic energy
with $B$, a positive definite inverse mass tensor. Applying the previous
results with the mapping $\mathbf{Q}\rightarrow\mathbf{P}$, $\mathbf{K}%
\rightarrow\mathbf{R}$ (and $i\nabla_{\mathbf{K}}\rightarrow-i\nabla
_{\mathbf{R}}$) we have $V(\mathbf{Q})\rightarrow B_{ij}P_{i}P_{j}/2$
and\textbf{ }we can check that the operator $T(\mathbf{R})$ corresponds to
$T(\mathbf{R})=\frac{p^{2}}{2m}+\varphi(\mathbf{R,r})$. We assume that its
eigenvalues $E_{n}(\mathbf{R})$ are known (or equivalently the matrix $U_{0}$
diagonalizing $T$). This eigenvalues are the energy levels of the fast system
for given position $\mathbf{R}$ of the slow one, such that the matrix elements
in this representation reads $\left\langle m\right\vert T(\mathbf{R}%
)\left\vert n\right\rangle =\delta_{m,n}E_{n}(\mathbf{R}).$ In this
representation the Hamiltonian of the slow part is non-diagonal but we can now
directly apply formula Eq. (\ref{epsilonh2}) to obtain the following
eigenvalues of $H$ to order $\hbar^{2}$ in terms of the slow variables
(assuming a non-degenerate spectrum for the fast system) :%

\begin{equation}
\varepsilon_{n}\left(  \mathbf{p}_{n},\mathbf{R}\right)  =\frac{1}{2}%
B_{ij}p_{n}^{i}p_{n}^{j}+\hbar^{2}\Phi_{n}+E_{n}(\mathbf{R})
\end{equation}
where $\mathbf{p}_{n}=\mathbf{P}-i\hbar\left\langle n\right.  \left\vert
\nabla_{\mathbf{R}}n\right\rangle $ is the gauge invariant momentum of the
slow system. The scalar potential Eq. $\left(  \ref{scalar}\right)  $ then
becomes
\begin{equation}
\Phi_{n}(\mathbf{R,P})=\frac{G_{n}^{ij}(\mathbf{R})}{2}B_{ij}\mathbf{+}%
M_{n}^{ij}(\mathbf{R})B_{il}B_{jk}P^{l}P^{k}%
\end{equation}
with the quantum metric tensor $G_{n}^{ij}(\mathbf{R})=\operatorname{Re}%
\sum_{m\neq n}\left\langle \partial_{i}n\right.  \left\vert m\right\rangle
\left\langle m\right.  \left\vert \partial_{j}n\right\rangle $ and $M_{n}%
^{ij}(\mathbf{R})=\operatorname{Re}\sum_{m\neq n}\frac{\left\langle
\partial_{i}n\right.  \left\vert m\right\rangle \left\langle m\right.
\left\vert \partial_{j}n\right\rangle }{\varepsilon_{0,m}-\varepsilon_{0,n}}.$
The term $G_{n}^{ij}(\mathbf{R})B_{ij}$ is the usual part of the scalar
potential discussed in several circumstances \cite{BERRY2}\cite{BERRY3}%
\cite{STERN}, whereas the term $M_{n}^{ij}(\mathbf{R})B_{il}B_{jk}P^{l}P^{k}$
was found in \cite{LITTLEJOHN}. Here we see that the Born-Oppenheimer theory
can be obtained straightforwardly from our Hamiltonian diagonalization to
order $\hbar^{2}.$ In the same manner from Eq.$\left(  \ref{EQMVT}\right)
$\ we immediately get the Born-Oppenheimer equations of motion $\overset
{\cdot}{\mathbf{p}}_{n}=-\nabla_{\mathbf{R}}E_{n}-\frac{\hbar}{2}%
(\overset{\cdot}{\mathbf{R}}\times\mathbf{\Theta}_{n}-\mathbf{\Theta}%
_{n}\times\overset{\cdot}{\mathbf{R}})-\hbar^{2}\nabla_{\mathbf{R}}\Phi_{n}$
with $\overset{\cdot}{R}_{i}=B_{ij}p_{n}^{j}$. Similar equations of motion for
a classical system consisting of a classical magnetic moment interacting with
an inhomogeneous magnetic field \cite{BERRY3}\cite{STERN} were studied in
details. It was found that the Lorentz force results from a slight
misalignment of the magnetic moment relative to the magnetic field. This
corresponds to the semi-classical approximation. The electric force is a time
average of a strong oscillatory force induced by the precession of the
magnetic moment. This is a kind of zitterbewegung effect.

\section{Particle in a linear external potential.}

Another interesting relevant situation concerns a particle in a linear
potential exemplified here by a Bloch electron in a constant external electric
field (see also \cite{HANSSEN} and for the first semi-classical treatment see
\cite{NIU}). Consider $H=H_{0}(\mathbf{P,R})+\varphi(\mathbf{R})$ with $H_{0}$
the energy of a particle in a periodic potential and $\varphi(\mathbf{R}%
)=-e\mathbf{E.R}$ the external electric perturbation (and $e<0$ the charge).
Again assuming that one knows $U_{0}$ diagonalizing $H_{0}$ we then have
$U_{0}H_{0}U_{0}^{+}=\varepsilon_{0,n}\left(  \mathbf{k}\right)  $ with
$\varepsilon_{0,n}\left(  \mathbf{k}\right)  $ the $n$-th energy band and
$\mathbf{k}$ the pseudo-momentum in the absence of the external field. Or, we
can also write $\left\langle u_{m}(\mathbf{k})\right\vert H_{0}(\mathbf{P}%
,\mathbf{R})\left\vert u_{n}(\mathbf{k})\right\rangle =\delta_{m,n}%
\varepsilon_{0,n}\left(  \mathbf{k}\right)  $ with $\left\vert u_{n}%
(\mathbf{k})\right\rangle $ the periodic part of the Bloch wave function (for
a more detailed discussion see also \cite{PIERRE2}).

Then, for the determination of the full eigenvalues of $H$ we use formula
(\ref{epsilonh2}) with the mapping $\mathbf{Q}\rightarrow\mathbf{R,}$ so that
the scalar gauge potential reduces to $\Phi_{n}(\mathbf{k})=e^{2}T_{n}%
^{ij}E_{i}E_{j}$, and the energy eigenvalues are
\begin{equation}
\varepsilon_{n}=\varepsilon_{0,n}\left(  \mathbf{k}\right)  -e\mathbf{E.r_{n}%
+}e^{2}\hbar^{2}M_{n}^{ij}\left(  \mathbf{k}\right)  E_{i}E_{j} \label{TIJ}%
\end{equation}
with the covariant coordinate $\mathbf{r}_{n}=\mathbf{R}+i\hbar\left\langle
u_{n}\right.  \left\vert \nabla_{\mathbf{k}}u_{n}\right\rangle $ and
$M_{n}^{ij}(\mathbf{k})=\operatorname{Re}\sum_{m\neq n}\frac{\left\langle
\partial_{i}u_{n}\right.  \left\vert u_{m}\right\rangle \left\langle
u_{m}\right.  \left\vert \partial_{j}u_{n}\right\rangle }{\varepsilon
_{0,m}-\varepsilon_{0,n}}$. Introducing the "magnetic field" $\mathbf{\omega
}_{n}\left(  \mathbf{k}\right)  =\frac{\hbar}{eE^{2}}\mathbf{E\times}%
\nabla_{\mathbf{k}}\Phi_{n}$ and $\mathbf{\chi}_{n}\left(  \mathbf{k}\right)
=\frac{1}{eE^{2}}\mathbf{E.}\nabla_{\mathbf{k}}\Phi_{n}$ the equations of
motion are%
\[
\overset{\cdot}{\mathbf{r}}_{n}=\nabla_{\mathbf{k}}\varepsilon_{0,n}%
-\frac{\hbar}{2}(\overset{\cdot}{\mathbf{k}}\times\Omega_{n}-\Omega_{n}%
\times\overset{\cdot}{\mathbf{k}})+\hbar^{2}\mathbf{\chi}_{n}\overset{\cdot
}{\mathbf{k}},\ \ \overset{\cdot}{\mathbf{k}}=e\mathbf{E}%
\]
where $\Omega_{n}\left(  \mathbf{k}\right)  =\mathbf{\Theta}_{n}\left(
\mathbf{k}\right)  +\hbar\mathbf{\omega}_{n}\left(  \mathbf{k}\right)  .$ This
shows that $\Phi_{n}$ contributes to the Lorentz term $\hbar\overset{\cdot
}{\mathbf{k}}\times\Omega_{n}$ known as the anomalous velocity which is
orthogonal to the applied electric field. This anomalous velocity is at the
center of many\ recent experimental and theoretical works and led to the
discovery the magnetic-type monopole in solids \cite{FANG}. The scalar
potential $\Phi_{n}$ contributes also to the velocity in the direction of
$\mathbf{E,}$ through the term $\hbar^{2}\mathbf{\chi}_{n}\overset{\cdot
}{\mathbf{k}}.$

\section{Beyond the Berry phase.}

The linear potential case has another interest. It allows us to also consider
the fast system to derive the Berry phase in a different way and to get a
correction term to the phase of the wave function.

\subsection{General results}

Indeed, consider a time dependent Hamiltonian $H(t)$ and introduce the
differential operator $D=H\left(  t\right)  -P_{0}$ where $P_{0}\equiv
i\hbar\partial/\partial t$ is the conjugate of time which is treated formally
as an operator such that $\left[  P_{0},t\right]  =i\hbar$. The time
dependence is due to the time evolution of some parameters $x(t)$\ describing
the environment. Assuming that we know $\varepsilon_{n}\left(  t\right)  $ and
$n(t)$ the instantaneous eigenvalues and eigenstates of $H$, i.e.,
$H(t)\left\vert n(t)\right\rangle =\varepsilon_{n}\left(  t\right)  \left\vert
n(t)\right\rangle ,$ then $D$ is a non-diagonal matrix in this representation
because of the presence of the time derivative operator $P_{0}$. To transform
the system of differential equations (Schr\"{o}dinger equation) $D\left\vert
\Psi(t)\right\rangle =0$, which couples all components of $\left\vert
\Psi(t)\right\rangle $ into a decoupled set of differential equations, we
introduce a unitary transformation $\left\vert \Psi^{\prime}\left(  t\right)
\right\rangle =U(t)\left\vert \Psi\left(  t\right)  \right\rangle $ such that
$U(t)D\left(  t,P_{0}\right)  U^{+}(t)=\widetilde{\Lambda}(t,P_{0})$ is a
diagonal differential operator and $\widetilde{\Lambda}(t,P_{0})\left\vert
\Psi^{\prime}\left(  t\right)  \right\rangle =0$. Therefore the time evolution
is given by $\left\vert \Psi^{\prime}\left(  t\right)  \right\rangle
=e^{\frac{-i}{\hbar}\int\nolimits_{0}^{t}\Lambda(t)dt}\left\vert \Psi^{\prime
}\left(  0\right)  \right\rangle .$ Since $\Lambda(t)=\widetilde{\Lambda
}(t)+P_{0}$ is diagonal, no time ordered product is required. Returning back
to the initial state we have%
\begin{equation}
\left\vert \Psi\left(  t\right)  \right\rangle =U^{+}(t)e^{\frac{-i}{\hbar
}\int\nolimits_{0}^{t}\Lambda(t)dt}U(0)\left\vert \Psi\left(  0\right)
\right\rangle \label{PHI}%
\end{equation}
A system prepared in a state $\left\vert \Lambda_{n}(0)\right\rangle $ which
is an eigenstate of $D$, i.e., $D(0)\left\vert \Lambda_{n}(0)\right\rangle
=\widetilde{\Lambda}_{n}(0)\left\vert \Lambda_{n}(0)\right\rangle $ will
evolve with $\Lambda(t)$ and thus stays in the instantaneous eigenstates
$\left\vert \Lambda_{n}(t)\right\rangle $ of $D(t)$ (for simplicity we assume
non degenerate eigenvalues). In this case the wave function becomes
$\left\vert \Psi\left(  t\right)  \right\rangle =e^{\frac{-i}{\hbar}%
\int\nolimits_{0}^{t}\Lambda_{n}(t)dt}\left\vert \Lambda_{n}(t)\right\rangle
$. Since eigenstates of $D$ instead of $H$ are considered, the time evolution
Eq. $\left(  \ref{PHI}\right)  $ is exact and thus valid for adiabatic as well
as for nonadiabatic evolution. Therefore for a periodic motion of period $T$,
such that $\left\vert \Lambda_{n}(T)\right\rangle =\left\vert \Lambda
_{n}(0)\right\rangle $ (single valued eigenstates), we have, if $\left\vert
\Psi\left(  0\right)  \right\rangle =\left\vert \Lambda_{n}(0)\right\rangle $
\begin{equation}
\left\vert \Psi\left(  T\right)  \right\rangle =e^{\frac{-i}{\hbar}%
\int\nolimits_{0}^{T}\varepsilon_{n}\left(  t\right)  dt}e^{-i\beta_{n}%
}\left\vert \Psi\left(  0\right)  \right\rangle \label{phitt}%
\end{equation}
with $\beta_{n}=\frac{1}{\hbar}\int\nolimits_{0}^{t}\Lambda_{n}(t)dt-\frac
{1}{\hbar}\int\nolimits_{0}^{T}\varepsilon_{n}\left(  t\right)  dt$ the exact
geometric phase of the system without any approximation which corresponds to
the Aharonov-Anandan phase \cite{AA}. Indeed, these authors extended the
notion of Berry geometric phase for cyclic adiabatic evolutions to
nonadiabatic cyclic evolutions (see also \cite{FERNANDEZ} and references therein).

But, in general we need an approximation scheme for the diagonalization of $D$
and we will use the expansion to order $\hbar^{2}.$ The problem of finding
$\Lambda_{n}=\widetilde{\Lambda}_{n}(t)+P_{0}$ is formally equivalent to the
Bloch electron example discussed above with $K\rightarrow t,$ $Q\equiv
R\rightarrow$ $P_{0}$ and $eE=1$. We obtain from Eq.$\left(  \ref{TIJ}\right)
$%

\begin{equation}
\Lambda_{n}=\varepsilon_{n}\left(  t\right)  -i\hbar\left\langle n\left\vert
\overset{\cdot}{n}\right\rangle \right.  \mathbf{+}\hbar^{2}M_{n}\left(
t\right)  \label{Dn}%
\end{equation}
with $M_{n}\left(  t\right)  =\operatorname{Re}\sum_{m\neq n}\frac
{\left\langle \overset{\cdot}{n}\right.  \left\vert m\right\rangle
\left\langle m\right.  \left\vert \overset{\cdot}{n}\right\rangle
}{\varepsilon_{m}-\varepsilon_{n}}.$ Therefore for a periodic motion of period
$T$, Eq. $\left(  \ref{phitt}\right)  $ becomes
\begin{equation}
\left\vert \Psi\left(  T\right)  \right\rangle =e^{-i\gamma_{n}}\left\vert
\Psi\left(  0\right)  \right\rangle \label{phiT}%
\end{equation}
with%
\[
\gamma_{n}=\frac{1}{\hbar}\int\nolimits_{0}^{T}\varepsilon_{n}(t)dt+i\int
_{0}^{T}\left\langle n\left\vert \overset{\cdot}{n}\right\rangle dt\right.
\mathbf{+}\hbar\int_{0}^{T}M_{n}\left(  t\right)  dt
\]
The phase $\gamma_{n}$\ appears as an expansion in power of $\hbar$. The first
term is the usual dynamical phase and the second one the geometric Berry phase
independent of $\hbar$\ and of the velocity of parameters $\overset{\cdot}%
{x}(t)$. The additional phase $\hbar\int_{0}^{T}M_{n}\left(  t\right)  dt$\ of
order $\hbar$\ is apparently non-geometric as it depends on $\overset{\cdot
}{x}$. It cancels in the infinitely slow $\overset{\cdot}{x}\rightarrow
0$\ adiabatic regime, which thus coincides with the semiclassical
approximation.\textbf{\ }Nevertheless, it is worth noticing that the higher
order phase corrections can also be seen as geometric in the Aharonov-Anandan
meaning \cite{BLIOKH}\cite{Rigolin}, since $\int_{0}^{T}M_{n}\left(  t\right)
dt$ can also be presented as a contour integral in a generalized parameter
space \cite{BLIOKH}.

Quantitatively, if the system is prepared in an eigenstate $\left\vert
n(0)\right\rangle $ of $H(0)$, then $\left\vert \Psi(t)\right\rangle $ is
given by Eq. $\left(  \ref{PHI}\right)  $ with $U=U_{0}+\hbar U_{1}U_{0}$
where $U_{1}\left(  t\right)  _{mn}=i\frac{\left(  1-\delta_{mn}\right)
\left\langle m\left\vert \overset{\cdot}{n}\right\rangle \right.
}{\varepsilon_{m}\left(  t\right)  -\varepsilon_{n}\left(  t\right)  }$, so
that we have the following expansion up to order $\hbar$:%
\begin{align*}
\left\vert \Psi(t)\right\rangle  &  =e^{\frac{-i}{\hbar}\int\nolimits_{0}%
^{t}\Lambda_{n}(t)dt}\left\vert n(t)\right\rangle +\hbar\sum_{m\neq n}\left(
e^{\frac{-i}{\hbar}\int\nolimits_{0}^{t}\Lambda_{n}(t)dt}A_{mn}(t)\right. \\
&  -\left.  e^{\frac{-i}{\hbar}\int\nolimits_{0}^{t}\Lambda_{m}(t)dt}%
A_{mn}(0)\right)  \left\vert m(t)\right\rangle +O(\hbar^{2})
\end{align*}
The magnitude of transitions is then controlled by the term $A_{mn}%
=\frac{i\left\langle m(t)\left\vert \overset{\cdot}{n}(t)\right\rangle
\right.  }{\varepsilon_{n}\left(  t\right)  -\varepsilon_{m}\left(  t\right)
}$ which is neglected in the strict adiabatic limit $\left\langle
m(t)\left\vert \overset{\cdot}{n}(t)\right\rangle \right.  \rightarrow0.$

In principle deviation from adiabaticity at the semi-classical level could be
measured by interferometry. Consider a periodic two states system, and write
the initial state in the eigenbase $\left\vert n(0)\right\rangle =\left\vert
\Lambda_{n}(0)\right\rangle +\hbar A_{mn}(0)\left\vert \Lambda_{m}%
(0)\right\rangle .$ Then, after one cycle $\left\vert \Psi\left(  T\right)
\right\rangle =e^{-i\gamma_{n}}\left\vert \Lambda_{n}(0)\right\rangle +\hbar
e^{-i\gamma_{m}}A_{mn}(0)\left\vert \Lambda_{m}(0)\right\rangle .$ For an
observable $O$ which does not commute with $H$ one will find in the average
$\left\langle \Psi\left(  T\right)  \right\vert O\left\vert \Psi\left(
T\right)  \right\rangle $ an interference term $2\hbar\operatorname{Re}\left(
A_{mn}(0)\left\langle \Lambda_{n}(0)\right\vert O\left\vert \Lambda
_{m}(0)\right\rangle e^{-i\left(  \gamma_{n}-\gamma_{m}\right)  }\right)  $
which is formally equivalent to the zitterbewegung of Dirac particles.

Note that $\left\vert \Psi(t)\right\rangle $ is normalized to unity at order
$\hbar$ only; i.e. $\left\langle \Psi(t)\right.  \left\vert \Psi
(t)\right\rangle =1+O(\hbar^{2})$ so that a normalization at a higher order
needs an expansion of $U$ to the same order. To second order, the
diagonalizing matrix $U_{2}\left(  t\right)  $ is given by the procedure
described in \cite{PIERRE1} leading to $U_{2}\left(  t\right)  _{mn}%
=-\sum_{k\neq n,m}\frac{<m\mid\overset{\cdot}{k}><k\mid\overset{\cdot}{n}%
>}{\left(  \varepsilon_{m}\left(  t\right)  -\varepsilon_{k}\left(  t\right)
\right)  \left(  \varepsilon_{k}\left(  t\right)  -\varepsilon_{n}\left(
t\right)  \right)  }$ so that the wave function reads%
\begin{align}
\left\vert \Psi(t)\right\rangle  &  =\left(  e^{\frac{-i}{\hbar}%
\int\nolimits_{0}^{t}\Lambda_{n}(t)dt}-\hbar^{2}\sum_{m\neq n}e^{\frac
{-i}{\hbar}\int\nolimits_{0}^{t}\Lambda_{m}(t)dt}A_{nm}(t)A_{mn}(0)+\hbar
^{2}e^{\frac{-i}{\hbar}\int\nolimits_{0}^{t}\Lambda_{n}(t)dt}\left(  U_{2}%
^{+}\left(  t\right)  _{nn}+U_{2}\left(  0\right)  _{nn}\right)  \right)
\left\vert n(t)\right\rangle \nonumber\\
&  \left(  \hbar\sum_{m\neq n}\left(  e^{\frac{-i}{\hbar}\int\nolimits_{0}%
^{t}\Lambda_{n}(t)dt}A_{mn}(t)-e^{\frac{-i}{\hbar}\int\nolimits_{0}^{t}%
\Lambda_{m}(t)dt}A_{mn}(0)\right)  +\hbar^{2}\sum_{m\neq n}\left(
e^{\frac{-i}{\hbar}\int\nolimits_{0}^{t}\Lambda_{n}(t)dt}U_{2}^{+}\left(
t\right)  _{mn}+e^{\frac{-i}{\hbar}\int\nolimits_{0}^{t}\Lambda_{m}(t)dt}%
U_{2}\left(  0\right)  _{mn}\right)  \right. \nonumber\\
&  \left.  -\hbar^{2}\sum_{p,m\neq n}e^{\frac{-i}{\hbar}\int\nolimits_{0}%
^{t}\Lambda_{p}(t)dt}A_{mp}(t)A_{pn}(0)\right)  \left\vert m(t)\right\rangle
\label{phicube}%
\end{align}
From this result we can compute the so called fidelity defined as $\left\vert
\left\langle \Psi_{ad}(t)\right\vert \left\vert \Psi(t)\right\rangle
\right\vert ^{2}$ where $\left\vert \Psi_{ad}(t)\right\rangle $ correspond to
the wave function in the adiabatic limit.\ Using Eq. $\left(  \ref{phicube}%
\right)  $ we obtain%

\begin{align}
\left\vert \left\langle \Psi_{ad}(t)\right\vert \left\vert \Psi
(t)\right\rangle \right\vert ^{2} &  =1-\hbar^{2}\sum_{k\neq n}\left(
\frac{\mid<n\mid\overset{\cdot}{k}>\mid^{2}}{\left(  \varepsilon_{n}\left(
t\right)  -\varepsilon_{k}\left(  t\right)  \right)  ^{2}}+\frac{\mid
<n\mid\overset{\cdot}{k}>\mid_{0}^{2}}{\left(  \varepsilon_{n}\left(
0\right)  -\varepsilon_{k}\left(  0\right)  \right)  ^{2}}\right)  \nonumber\\
&  -\hbar^{2}\sum_{k\neq n}\frac{e^{-i(\gamma_{k}-\gamma_{n})}<n\mid
\overset{\cdot}{k}><k\mid\overset{\cdot}{n}>_{0}+e^{i(\gamma_{k}-\gamma_{n}%
)}\left(  <n\mid\overset{\cdot}{k}><k\mid\overset{\cdot}{n}>_{0}\right)
^{\ast}}{\left(  \varepsilon_{n}\left(  t\right)  -\varepsilon_{k}\left(
t\right)  \right)  \left(  \varepsilon_{n}\left(  0\right)  -\varepsilon
_{k}\left(  0\right)  \right)  }\label{fidelity}%
\end{align}
with $e^{-i(\gamma_{k}-\gamma_{n})}=e^{-i\int\left[  \frac{\varepsilon
_{k}-\varepsilon_{n}}{\hbar}+(\phi_{k}-\phi_{n})+\hbar(M_{k}-M_{n})\right]
dt}$ with $\phi_{n}=\left\langle n\left\vert \overset{\cdot}{n}\right\rangle
\right.  .$

\subsection{Example}

As a physical illustration let consider the Hamiltonian%

\[
H=\mathbf{B}\left(  t\right)  \cdot\mathbf{\sigma}%
\]
corresponding to the paradigmatic example of a spin coupled to a magnetic
field. The eigenvalues of $D=H\left(  t\right)  -i\hbar\partial/\partial t$ at
order $\hbar^{2}$ are given by Eq. $\left(  \ref{Dn}\right)  $%

\begin{equation}
\Lambda_{l}=\varepsilon_{l}\left(  t\right)  +\hbar a_{l}^{0}\mathbf{+}%
\hbar^{2}M_{l}\left(  t\right)
\end{equation}
The Hamiltonian being diagonalized by the matrix $U_{0}=\frac{(\mathbf{n\cdot
\sigma+}\sigma_{z})}{\sqrt{2\left(  1+n_{z}\right)  }}$ with $\mathbf{n}%
=\mathbf{B/}B,$ we have the two level eigenvalues $\varepsilon_{l}\left(
t\right)  =B(t)\left(  \sigma_{z}\right)  _{ll}$ with $l=1,2,$ and the matrix
$A^{0}=-i\hbar U_{0}\frac{\partial}{\partial t}U_{0}^{+}$ is thus given
\[
A^{0}=-\hbar\left(  \frac{\left(  \left(  \mathbf{n+k}\right)  \mathbf{\times
}\overset{\cdot}{\mathbf{n}}\right)  .\mathbf{\sigma}}{2\left(  1+n_{z}%
\right)  }\right)
\]
As a consequence the scalar gauge potential $a_{l}^{0}=-i<l\mid\overset{\cdot
}{l}>=-i\hbar\left(  U_{0}\frac{\partial}{\partial t}U_{0}^{+}\right)  _{ll}$
reads $a_{l}^{0}=-\frac{\hbar\left(  \left(  \mathbf{n+k}\right)
\mathbf{\times}\overset{\cdot}{\mathbf{n}}\right)  \cdot\mathbf{k}}{2\left(
1+n_{z}\right)  }\left(  \sigma_{z}\right)  _{ll}$ and $M_{l}\left(  t\right)
=M_{l}^{00}=\frac{1}{2}\sum\nolimits_{m\neq l}(\frac{(A^{0})_{lm}(A^{0})_{ml}%
}{\varepsilon_{,m}-\varepsilon_{0,l}}+h.c.)$ can therefore be written
\begin{equation}
M_{l}=-\frac{2\left(  1+n_{z}\right)  \overset{\cdot}{n}^{2}-\overset{\cdot
}{n_{z}}^{2}-\left(  \left(  \mathbf{n\times}\overset{\cdot}{\mathbf{n}%
}\right)  \cdot\mathbf{k}\right)  ^{2}}{8\left(  1+n_{z}\right)
^{2}\varepsilon_{l}(t)}\label{TT}%
\end{equation}
The Berry phase can thus be cast in the form
\begin{equation}
\phi_{l,B}=-\left(  \sigma_{z}\right)  _{ll}\int_{0}^{T}dt\frac{\left(
\mathbf{n}\times\overset{\cdot}{\mathbf{n}}\right)  .\mathbf{k}}{2\left(
1+n_{z}\right)  }\label{phiB}%
\end{equation}
It is interesting to introduce the Euler angles $\mathbf{n=(}\sin\theta
\cos\varphi,\sin\theta\sin\varphi,\cos\theta\mathbf{),}$ so that the Berry
phase reads $\phi_{l,B}=\left(  \sigma_{z}\right)  _{ll}\frac{1}{2}\int
_{L}\left(  1-\cos\theta\right)  d\varphi$ and the correction of higher order
to the phase $\gamma_{l}$ is thus
\begin{equation}
\hbar\int_{0}^{T}M_{l}\left(  t\right)  dt=\frac{\hbar}{8B}\int_{0}^{T}%
(\sin^{2}\theta\overset{\cdot}{\varphi}^{2}+\overset{\cdot}{\theta}^{2})dt.
\end{equation}
We retrieve the result of \cite{BLIOKH} for $\varphi=0.$ Note that, it was
shown in \cite{BLIOKH}, this higher order contribution to the geometric phase
is a part of the non-adiabatic Aharonov-Anandan geometric phase.

Let compute the fidelity in the simple case $\theta=const$ and $\varphi=\omega
t.$ A direct application of Eq. $\left(  \ref{fidelity}\right)  $ gives
\begin{equation}
\left\vert \left\langle \Psi_{ad}(t)\right\vert \left\vert \Psi
(t)\right\rangle \right\vert ^{2}=1-\frac{\omega^{2}\sin^{2}\theta}%
{4\omega_{0}^{2}}\sin^{2}\left(  \omega_{0}-\omega\right)  t+...
\end{equation}
where we introduced the notation $B=\hbar\omega_{0}.$ The usual criteria for
adiabaticity is given by the condition $\frac{\sin^{2}\theta\omega^{2}%
}{4\omega_{0}^{2}}<<1.$ At the resonance $\frac{\omega_{0}}{\omega}=1$
fidelity is always one at this order of the expansion.

\section{Dirac particle in an external potential.}

We will now show that our formalism can also be used for relativistic Dirac
particles, which are usually treated with the Foldy Wouthuysen approach
\cite{FOLDY}. The semiclassical treatment of the problem was first considered
in \cite{ALAIN} and \cite{BLIOKH3}. The direct second order in $\hbar$
diagonalization being done in \cite{PIERRE1}, we can now fully appreciate how
simple and straightforward is the application of the formalism presented here.

The Hamiltonian is (with $c=1$)%
\begin{equation}
H=\mathbf{\alpha.p}+\beta m+V\left(  \mathbf{R}\right)
\end{equation}
where $\mathbf{\alpha}$ and $\beta$ are the usual $\left(  4\times4\right)  $
Dirac matrices and $V\left(  \mathbf{R}\right)  $ is the external potential.
The matrix diagonalizing the free part of the Hamiltonian $U_{0}\left(
\mathbf{\alpha.p}+\beta m\right)  U_{0}^{+}=\beta E$ with $E=\sqrt
{\mathbf{p}^{2}+m^{2}}$ is the usual Foldy Wouthuysen unitary transformation
$U_{0}=\frac{E+m+\beta\mathbf{\alpha P}}{\left(  2E\left(  E+m\right)
\right)  ^{1/2}}.$ For the Dirac particles we have two energy subspaces
$\varepsilon_{\pm}$ of dimension 2 corresponding to the positive and negative
energy. Now with the correspondence $\mathbf{Q}\rightarrow\mathbf{R},$
$\mathbf{K}\rightarrow\mathbf{p}$ and formula Eq. $\left(  \ref{epsilonh2}%
\right)  ,$ one easily sees that the diagonal matrix can be written%
\begin{equation}
\varepsilon\left(  \mathbf{p,r}\right)  =\beta E\left(  \mathbf{p}\right)
+V\left(  \mathbf{r}\right)  +\beta\Phi\label{EDirac}%
\end{equation}
The position operator is given by the ($4\times4$) matrix $\mathbf{r=R+}%
\frac{\hbar\mathbf{p\times\Sigma}}{2E\left(  E+m\right)  }$ with
$\mathbf{\Sigma=1\otimes\sigma}$ where $\mathbf{\sigma}$ are the Pauli
matrices. The band index of the scalar potential has been transferred to the
matrix $\beta$, and we have $G^{ij}=\frac{1}{4E^{2}}g^{ij}$ and $M^{ij}%
=\frac{1}{8E^{3}}g^{ij}$ with the notation $g^{ij}=\delta^{ij}-\frac
{p^{i}p^{j}}{E^{2}},$ so that finally we can write
\begin{equation}
\Phi=\frac{\hbar^{2}}{8E^{2}}g^{ij}(\mathbf{\partial}_{i}\mathbf{\partial}%
_{j}V\mathbf{+}\frac{1}{E}\mathbf{\partial}_{i}V\partial_{j}V)
\label{phidirac}%
\end{equation}
Note the presence of a new term not presented in other considerations of the
Dirac equation (see for instance \cite{Zawadzki}), which is nonlinear in
scalar potential and which stems from the new tensor $M^{ij}.$ If for central
potential one can neglect this new contribution $\frac{1}{E}\mathbf{\partial
}_{i}V\partial_{j}V$, this is not always true and for some potentials both
terms in Eq.\ $\left(  \ref{phidirac}\right)  $ can be of the same magnitude.
In fact for constant electric field $V=-e\mathcal{E}\mathbf{.R}$, the first
term vanishes and $\Phi=\frac{e^{2}\hbar^{2}}{8E^{2}}g^{ij}\mathcal{E}%
_{i}\mathcal{E}_{j}$.

In the non-relativistic limit $\mathbf{p}<<m$, $\Phi$ becomes $\Phi
\approx\frac{\hbar^{2}}{8m^{2}}\left(  \mathbf{\Delta}V\mathbf{+}\frac{1}%
{m}(\mathbf{\nabla}V)^{2})\right)  +O(\hbar^{2}p^{2}/m^{4})$ which gives two
contributions. The first one is the usual Darwin term $\frac{\hbar^{2}}%
{8m^{2}}\mathbf{\Delta}V$ traditionally obtained as the result of the Foldy
Wouthuysen transformation expanded in power of $1/m.$ The second term
$\frac{\hbar^{2}}{8m^{3}}\left(  \mathbf{\nabla}V\right)  ^{2}$ of higher
order in $1/m$ is usually not considered in the Foldy Wouthuysen approach. It
is also interesting to note that the external potential in the non
relativistic limit can be expanded as $V\left(  \mathbf{r}\right)  \approx
V\left(  \mathbf{R}\right)  +\frac{\hbar}{4m^{2}}\mathbf{\Sigma\mathbf{.}%
}\left(  \mathbf{\nabla}V\times\mathbf{p}\right)  +O(\hbar^{2}p^{2}/m^{4})$
where $\frac{\hbar}{4m^{2}}\mathbf{\Sigma\mathbf{.}}\left(  \mathbf{\nabla
}V\times\mathbf{p}\right)  $ is the spin-orbit coupling term. Therefore the
Hamiltonian can be approximated as%

\begin{align}
\varepsilon &  \approx\beta\left(  m+\frac{\mathbf{P}^{2}}{2m}-\frac
{\mathbf{P}^{4}}{8m^{3}}\right)  +V\left(  \mathbf{R}\right)  +\frac{\hbar
}{4m^{2}}\mathbf{\Sigma\mathbf{.}}\left(  \mathbf{\nabla}V\times
\mathbf{p}\right) \nonumber\\
&  \mathbf{+}\frac{\hbar^{2}}{8m^{2}}\beta\left(  \mathbf{\Delta}%
V\mathbf{+}\frac{1}{m}(\mathbf{\nabla}V)^{2}\right)  \label{HM}%
\end{align}
A Born-Oppenheimer treatment of the Dirac equation where the spin is the fast
variable and the momentum the slow one has led to the same Hamiltonian
Eq.\ $\left(  \ref{HM}\right)  $ but without the scalar potential
\cite{MATHUR}. This corresponds to the semiclassical approximation. The
additional electric-type potential $\Phi$ is a consequence of transitions
between energy levels. This is in agreement with the usual interpretation of
the physical origin of the Darwin term, the zitterbewegung phenomenon, whereby
the electron does not move smoothly but instead undergoes extremely rapid
small-scale fluctuations due to an interference between positive and negative
energy states.

\section{Conclusion.}

In this paper we investigated the origin of quantum geometric phases, gauge
fields and forces beyond the adiabatic approximation for physical systems
displaying a separation of scales in terms of slow and fast degrees. In
particular we extended the notions of geometric magnetic and electric forces
discovered in studies of the Born-Oppenheimer approximation. Our approach is
very general and results found here have been straightforwardly applied to
several physical systems.

\textit{Acknowledgement}. We are grateful to Prof. M. V. Berry for having
drawn our attention to this subject. We are also grateful to a referee for
very helpful clarifications and comments which helped us to improve the
present paper.

\end{document}